\documentstyle[12pt]{article}%
\renewcommand{\title}[1]{\large\bf
     #1\bigskip\medskip\\}
\renewcommand{\author}[1]{\large #1\\ \smallskip}
\newcommand{\address}[1]{{\normalsize\it #1\\}\bigskip}

\newcommand{\be}{\begin{eqnarray}}
\newcommand{\ee}{\end{eqnarray}}
\newcommand{\hs}[1]{\hspace*{#1cm}}
\newcommand{\vs}[1]{\vspace*{#1cm}}
\newcommand{\no}{\nonumber}
\newcommand{\ig}[1]{\mbox{  }}
\newcommand{\wt}[6]{#1\mbox{\small
 $\left(\matrix{#5&#4\cr#2&#3\cr}\biggm|\mbox{$#6$}\right)$}}

\newcommand{\Km}[5]{{#1}\biggl(\!\matrix{&#3\vs{-0.3}\cr\!\!
  #2\hs{-0.3}\vs{-0.3}&\cr&#4}\!\!\biggm|\!\mbox{$#5$}\biggr)}
\newcommand{\Kp}[5]{{#1}\biggl(\matrix{#3\vs{-0.3}&\cr&
  \hs{-0.3}#2\vs{-0.3}\cr#4&}\!\!\biggm|\!\mbox{$#5$}\biggr)}

\newcommand{\Z}{\mbox{\sf Z\hspace*{-0.45em}Z}}

\begin{document}
\begin{center}
\title{Surface Critical Phenomena in\\ Interaction-Round-a-Face Models}
\author{Yu-Kui Zhou and Murray T. Batchelor}
\address{Department of Mathematics, School of Mathematical Sciences,\\
         The Australian National University, Canberra ACT 0200, Australia}

\begin{abstract} A general scheme has been proposed to study the
critical behaviour of integrable interaction-round-a-face models
with fixed boundary conditions. It has been shown that
the boundary crossing symmetry plays an important role in
determining the surface free energy. The
surface specific heat exponent can thus be obtained without
explicitly solving the reflection equations for the
boundary face weights.
For the restricted SOS $L$-state models of Andrews, Baxter and Forrester
the surface specific heat exponent is found to be
$\alpha_s=2-(L+1)/4$.
\end{abstract}
\end{center}

\clearpage
\setcounter{page}{1}

\subsection{Introduction}

Since Sklyanin's work \cite{Sklyanin}
integrable models with open boundary conditions have received
considerable attention (see, e.g.,
{\cite{MezNep,YuBa:95,Kulish,BOP:95} and references therein).
However, despite this flurry of activity, much of the exact surface
critical behaviour of the various models, both in quantum
field theory and in statistical mechanics, remains to be derived.
Our motivation here is to apply the boundary crossing symmetry
investigated by Ghoshal and Zamolodchikov in terms of quantum field theory
\cite{GhZa:94} to study surface critical phenomena in integrable
lattice models in statistical mechanics.

The particular models of interest are the interaction-round-a-face (IRF)
solid-on-solid (SOS) models. First among these is Baxter's SOS model
\cite{Baxter:73} which is intimately related to the
eight-vertex model \cite{Baxter:72}.
In this SOS model the heights round a given face are unrestricted.
Under restriction of the heights, this model becomes the $L$-state
restricted SOS model of Andrews, Baxter and Forrester (ABF) \cite{ABF:84}.
There has been increasing interest in studying these SOS models with
open boundary conditions \cite{BOP:95,Zhou:95b,AK:95}. In this case
the integrability governed by the usual star-triangle equation must
be supplemented by the reflection equations.
In \cite{Zhou:95b} the intertwining relation
between the vertex and face weights at the boundary has been studied and
thus, starting from the reflection equations of the eight-vertex
model, the IRF analogue of the reflection equations and the general
solutions of the boundary face weights have been formulated
using the boundary face-vertex intertwining relation.
Reflection equations for IRF models have also been written down
in \cite{Kulish,BOP:95,AK:95}.

In this paper we show that the boundary crossing symmetry
of the boundary face weights is enough to give the unitarity
relation for the surface free energy on a square lattice rotated by
$45^\circ$ with fixed boundary conditions. Thus we can extract
some surface critical behavior without knowing the explicit details of
the boundary face weights. The underlying models considered here
are the unrestricted SOS model and ABF restricted SOS models.
However, the discussion is applicable to other integrable IRF models.

In the next section we describe the models with open boundary
conditions. Instead of solving
the SOS reflection equations  we assume that the generic solutions
of the reflection equations define the boundary face weights,
 which satisfy the boundary crossing symmetry.
In section~\ref{sec3} we show how to construct
the unrestricted or restricted SOS models with fixed boundary
conditions from the models with the open boundary conditions.
This is done by taking the special open boundary face weights and
introducing alternating inhomogeneities such that the square lattice
is rotated by $45^\circ$.  In section~\ref{sec4} we present the
functional relations for whole fusion hierarchies of the models.
Ignoring the finite-size corrections to the transfer matrices we end
up with a unitarity relation determining both the bulk and surface free
energies. This relation is simplified by the boundary
crossing symmetry. Then in section~\ref{sec5} we study the surface
critical behavior of the models with fixed boundary conditions.
In particular, the surface critical exponent corresponding
to the specific heat does not rely on the details of
the boundary face weights. As a by-product, using the known
scaling relation, we also obtain the
correlation length exponent $\nu=(L+1)/4$ for the ABF model
with odd $L$.

\subsection{Models with open boundary conditions}
\setcounter{equation}{0}

The square lattice ABF models are
restricted solid-on-solid (RSOS) models with $L$ heights built
on the classical $A_L$ Dynkin diagram. The corresponding unrestricted
SOS model has been introduced by Baxter in the study of
the eight-vertex model \cite{Baxter:73}.
By using the intertwiners the
eight-vertex model can be transformed into the SOS model, which is
defined by the following nonzero face weights
\addtolength{\jot}{2mm}
\be
\wt Wa{a\pm 1}a{a\mp 1}u & = & {\vartheta_1(\lambda-u)
       \over \vartheta_1(\lambda)} \no \\
\wt W{a\pm 1}a{a\mp 1}au & = &
\left[\frac{\vartheta_1(w_{a-1})\vartheta_1(w_{a+1})}{
  \vartheta_1^2(w_a)}\right]^{1/2}
 \frac{\vartheta_1({u})}{\vartheta_1({\lambda})} \no \\
\wt W{a\pm 1}a{a\pm 1}au & = &
           \frac{\vartheta_1({w_a\pm u})}{\vartheta_1(w_a)}
\ee
where $w_a=a\lambda+w_0$. The height variable $a\in\Z$
with $w_0$ and $\lambda$ as free parameters.
The function $\vartheta_1({u})$
is a standard elliptic theta functions of nome $p$,
\be
\vartheta_1(u)=\vartheta_1(u,p)=2p^{1/8}\sin u\:\prod_{n=1}^{\infty}
\left(1-2p^{n}\cos 2u+p^{2n}\right)\left(1-p^{n}\right)\label{theta1}.
\ee
The above face weights satisfy the star-triangle equation
\be
\sum_g\wt Wabgfu\wt {W}fgdev\wt {W}gbcd{v\!-\!u} \no \\
=\sum_g\wt {W}fage{v\!-\!u}\wt {W}abcgv\wt Wgcdeu \label{YBE}
\ee
along with inversion/unitarity/crossing unitarity relations
\be
\sum_{g} \wt Wabgdu\wt Wgbcd{-u} =\rho(u)\delta_{a,c}
\ee
\be
\sum_{g} \wt Wdabg{\lambda-u} \wt
Wdgbc{\lambda+u} {\vartheta_1(w_{a})\vartheta_1(w_{g})
\over \vartheta_1(w_{d})\vartheta_1(w_{b})}=\rho(u)\delta_{a,c}
\ee
where $\rho(u)=\vartheta_1({\lambda-u})\vartheta_1({\lambda+u})/
\vartheta_1^2({\lambda})$.

The $A_L$ ABF RSOS models follow from the unrestricted SOS model on
setting $w_0=0$ and $\lambda={\pi\over L+1}$ with  $a=1,2,\cdots,L$,
where $L=3,4,\cdots$. In particular, the
Ising model is described by the $A_3$ model.

The integrable boundary face weights are represented by a triangular
face with three spins interacting round the face \cite{BOP:95,Zhou:95b},
\be
\Km {K}acbu=0 \hspace*{0.5cm}\mbox{unless $|a-b|=1$
   and $|a-c|=1$}.  \label{K}
\ee
They satisfy the reflection equation \cite{Kulish,Zhou:95b}
\be
&&\sum_{f,g}{\wt Wgcba{u-v}}{\Km {K}gcf{u;\xi}}{\wt
   Wdfga{u+v}}{\Km {K}dfe{v;\xi}}
  \no\\
&&\hspace{0.5cm} =\sum_{f,g} {\Km {K}bcf{v;\xi}}{\wt
   Wgfba{u+v}}{\Km {K}gfe{u;\xi}
     }{\wt Wdega{u-v}} \label{BYBE}
\ee
where generally we may have more arbitrary parameters
than one like $\xi$. The SOS analogue of boundary
crossing symmetry is given by
\be
\sum_{c}\sqrt{\vartheta_1(w_c)\over\vartheta_1(w_a)}
\wt Wabcd{2u+\lambda} \Km Kceb{u+\lambda} = {\vartheta_1(
 2u+2\lambda)\over \vartheta_1(\lambda)}    \Km Kaeb{-u}\;.
\ee
This relation is expected by applying the intertwining relation to
the boundary crossing symmetry of the eight-vertex model
\cite{GhZa:94,Zhou:95b}.

Following \cite{Zhou:95b} we define the row-to-row transfer matrix
$\mbox{\boldmath $T$}(u)$  with elements
\be
\langle\mbox{\boldmath $a$}|\mbox{\boldmath $T$}(u)
   |\mbox{\boldmath $b$}\rangle
  =\sum_{\{c_0,\cdots,c_N\}} \Kp {K_+}{c_0}{a_0}{b_0}u
  \Km {K_-}{c_N}{a_N}{b_N}u \times  \hspace{1cm} \nonumber\\
 \prod_{k=0}^{N-1} \biggl[\wt {W}{b_k}{b_{k+1}}{c_{k+1}}{c_k}{u-v_k}
 \wt {W}{c_{k+1}}{a_{k+1}}{a_k}{c_k}{u+v_k}\biggl]
   \; ,  \label{openT}
\ee
where $\mbox{\boldmath $a$}=\{a_0,a_1,\cdots,a_N\}$ and
$\mbox{\boldmath $b$}=\{b_0,b_1,\cdots,b_N\}$. Here $v_k$
are arbitrary parameters which play the role of inhomogeneities.
The right boundary face weights $K_-$ are given by
\be
\Km {K_-}{a}{c}{b}u =\Km {K}{a}{c}{b}{u;\xi_-}
\ee
and the left boundary face weights $K_+$ are given by
\be
\Kp {K_+}{a}{c}{b}u =\Km {K}{a}{c}{b}{-u+\lambda;
     \xi_+}\sqrt{\vartheta^2_1(w_{a})
     \over  \vartheta_1(w_{b})\vartheta_1(w_{c})}\;.
\ee
So defined, the transfer matrix forms a commuting family
\be
\left[\mbox{\boldmath $T$}(u)\; ,
  \; \mbox{\boldmath $T$}(v)\;\right]=0\;. \label{TT}
\ee
It follows that the above SOS and RSOS models with
open boundary conditions formulated
by the boundary face weights $K$ are integrable systems.

\subsection{ABF models with fixed boundary conditions}
\label{sec3}\setcounter{equation}{0}
\begin{figure}[t]
\begin{center}
\setlength{\unitlength}{0.00600in}%
\begin{picture}(390,258)(45,501)
\put(114,759){\line(-1,-1){ 39}}
\put( 75,720){\line( 1,-1){219}}
\put( 75,660){\line( 1,-1){159}}
\put( 75,600){\line( 1,-1){ 99}}
\put( 75,540){\line( 1,-1){ 36}}
\put( 75,660){\line( 1, 1){ 99}}
\put( 75,600){\line( 1, 1){156}}
\put( 75,540){\line( 1, 1){216}}
\put(219,756){\line( 1,-1){216}}
\put(435,540){\line(-1,-1){ 39}}
\put(276,759){\line( 1,-1){159}}
\put(435,600){\line(-1,-1){ 96}}
\put(156,759){\line( 1,-1){258}}
\put( 99,756){\line( 1,-1){252}}
\put(336,759){\line( 1,-1){ 99}}
\put(435,660){\line(-1,-1){156}}
\put(396,759){\line( 1,-1){ 39}}
\put(435,720){\line(-1,-1){216}}
\put(102,507){\line( 1, 1){252}}
\put(159,504){\line( 1, 1){255}}
\put(435,756){\line( 0,-1){249}}
\put( 45,750){\line( 0,-1){237}}
\put( 45,750){\line( 1,-1){ 30}}
\put( 75,720){\line(-1,-1){ 30}}
\put( 46,629){\line( 1,-1){ 30}}
\put( 76,599){\line(-1,-1){ 30}}
\put( 46,689){\line( 1,-1){ 30}}
\put( 76,659){\line(-1,-1){ 30}}
\put( 45,570){\line( 1,-1){ 30}}
\put( 75,540){\line(-1,-1){ 30}}
\multiput(96,714)(60,0){6}{\scriptsize$2u$}
\multiput(96,657)(60,0){6}{\scriptsize$2u$}
\multiput(96,597)(60,0){6}{\scriptsize$2u$}
\multiput(96,534)(60,0){6}{\scriptsize$2u$}
\multiput(126,684)(60,0){5}{\scriptsize$2u$}
\multiput(126,627)(60,0){5}{\scriptsize$2u$}
\multiput(126,567)(60,0){5}{\scriptsize$2u$}
\multiput(54,714)(0,-60){4}{\scriptsize$u$}
\multiput(420,687)(0,-60){4}{\scriptsize$u$}
\end{picture}
\caption{\small The rotated square lattice following from
  the choice (\ref{homo}) for the alternating inhomogeneities in
  the unrotated square lattice. Each face has spectral
  parameter $2u$ while the boundary faces have spectral parameter
  $u$.}
\end{center}
\bigskip
\begin{center}
\setlength{\unitlength}{0.00600in}%
\begin{picture}(360,258)(75,501)
\put(114,759){\line(-1,-1){ 39}}
\put( 75,720){\line( 1,-1){219}}
\put( 75,660){\line( 1,-1){159}}
\put( 75,600){\line( 1,-1){ 99}}
\put( 75,540){\line( 1,-1){ 36}}
\put( 75,660){\line( 1, 1){ 99}}
\put( 75,600){\line( 1, 1){156}}
\put( 75,540){\line( 1, 1){216}}
\put(219,756){\line( 1,-1){216}}
\put(435,540){\line(-1,-1){ 39}}
\put(276,759){\line( 1,-1){159}}
\put(435,600){\line(-1,-1){ 96}}
\put(156,759){\line( 1,-1){258}}
\put( 99,756){\line( 1,-1){252}}
\put(336,759){\line( 1,-1){ 99}}
\put(435,660){\line(-1,-1){156}}
\put(396,759){\line( 1,-1){ 39}}
\put(435,720){\line(-1,-1){216}}
\put(102,507){\line( 1, 1){252}}
\put(159,504){\line( 1, 1){255}}
\multiput(96,714)(60,0){6}{\scriptsize$2u$}
\multiput(96,657)(60,0){6}{\scriptsize$2u$}
\multiput(96,597)(60,0){6}{\scriptsize$2u$}
\multiput(126,684)(60,0){5}{\scriptsize$2u$}
\multiput(126,627)(60,0){5}{\scriptsize$2u$}
\multiput(126,567)(60,0){5}{\scriptsize$2u$}
\multiput(96,534)(60,0){6}{\scriptsize$2u$}
\multiput(438,714)(0,-60){4}{\small$d$}
\multiput(411,687)(0,-60){4}{\small$c$}
\multiput(61,717)(0,-60){4}{\small$b$}
\multiput(90,685)(0,-60){4}{\small$a$}
\end{picture}
\caption{\small The rotated square lattice with the boundary spins
$a,b,c,d$ fixed:
    $|a-b|=1$ and $|c-d|=1$. The boundary face weights are dropped after
    taking the free parameter $\xi=\xi_0$. }
\end{center}
\end{figure}
The square lattice rotated by $45^\circ$ is a natural geometry
to investigate surface critical phenomena.
The ABF model in this geometry has received particular attention
\cite{SaBa:89,Cardy:89}.
 From the perspective of exactly solved vertex models it is known
that the rotated geometry can be realised by
appropriate choices of the inhomogeneities in the unrotated
lattice \cite{DeDe:92,YuBa:95}.
Here we apply the same idea to the ABF model, however the
discussion is also valid for the unrestricted SOS model.
We show that the ABF model on the rotated lattice with fixed boundary
conditions follows from the ABF model on the unrotated
lattice with the corresponding open boundary conditions.

Let us consider the ABF models on the unrotated square lattice
with the row-to-row transfer matrix defined in (\ref{openT}).
For simplicity we consider only the diagonal $K$-matrices, or
alternatively, the boundary face weights (\ref{K}) vanish if $b\ne c$.
The particular choice of inhomogeneities
\be
v_k=(-1)^k u \label{homo}
\ee
leads to the desired orientation because of the property
\be
\wt Wabcd0 =\delta_{a,c}\;.
\ee
To see this it is most instructive to view the situation graphically
as was done for the vertex models in \cite{YuBa:95}. In this way
we arrive at the rotated lattice in Fig.~1, where the face weights
have been depicted graphically \vspace{-0.5cm} as
\be
\wt Wabcdu=\mbox{\setlength{\unitlength}{0.006500in}%
\begin{picture}(72,81)(95,711)
\put(135,750){\line(-1,-1){ 30}}
\put(105,720){\line( 1,-1){ 30}}
\put(135,690){\line( 1, 1){ 30}}
\put(165,720){\line(-1, 1){ 30}}
\put(171,717){\small$b$}
\put(95,714){\small$d$}
\put(131,678){\small$a$}
\put(131,753){\small$c$}
\put(132,714){\small$u$}
\end{picture}}
\ee

\noindent
The boundary face weights represented by the triangular faces
contain the free parameter $\xi$ which, following \cite{BOP:95}
can be chosen such that
$$\Km Kabb{u,\xi_0} =0 \hspace{1cm}
   \mbox{for $a=b+1$ or $a=b-1$.} $$
Thus the boundary face weights can be dropped under normalisation.
In this way we arrive at the ABF models on the rotated square lattice
with fixed boundary conditions, as depicted in Fig.~2.
This is the special geometry considered in \cite{SaBa:89,Cardy:89}
with the particular fixed boundary conditions corresponding to arrow
conservation at the boundaries in the vertex formulation.
Like their vertex model counterparts the
procedure shown here provides a possibility to exactly
solve an integrable SOS or RSOS model on the rotated square
lattice with fixed boundary conditions by first solving the
corresponding model on the unrotated square lattice.
Thus the coordinate, algebraic or analytic Bethe ansatz could play
an important role again.

\subsection{Fusion hierarchies and functional relations}
\label{sec4}\setcounter{equation}{0}

The fusion procedure for constructing integrable generalisations of
the SOS or ABF models has been described in
\cite{DJKMO:88} (see \cite{ZhPe:94} for a related work).
The fusion of integrable
open boundary face weights can be done in a similar manner to form
fusion hierarchies of new integrable models
defined via the fused weights \cite{MeNe:92,Zhou:95a,BOP:95,Zhou:95b}.
The functional relations of the fusion hierarchies
of the ABF models with diagonal open boundary conditions
($b=c$ in (\ref{K})) have been given in \cite{BOP:95}.
However, the transfer matrix defined in \cite{BOP:95}
is different to the one considered here, so it is a worthwhile
exercise to derive the functional relations for the fused transfer
matrix defined in (\ref{fopenT}) with
general boundary fused face weights based on (\ref{K}). The derivation is
in the same spirit as in \cite{BOP:95,Zhou:95b} and is to apply
the fusion procedure to the product of two transfer matrices.

For a square lattice the fusion can be completed separately in
the vertical and horizontal directions. The fused face weights
are expressed by $\wt {W_{(n,m)}}abcdu$ with the fusion levels
$n$ and $m$, respectively, in the vertical and horizontal directions
\cite{ZhPe:94}. The fused boundary face weights are given by
$\Km {K_-^{(n)}}acbu$ and $\Kp {K_+^{(n)}}acbu$ with fusion level
$n$. It has been shown that the
fused face weights satisfy the star-triangle equation and fused
boundary face weights satisfy the reflection equation
\cite{BOP:95,Zhou:95b}. Thus the fused models with open boundary
conditions are still integrable if the corresponding
fused transfer matrices $\mbox{\boldmath $T$}^{(m,n)}(u)$ are defined by
\be
\langle\mbox{\boldmath $a$}|\mbox{\boldmath $T$}^{(m,n)}(u)|
  \mbox{\boldmath $b$}\rangle =\sum_{\{c_0,\cdots,c_N\}}\Kp {K_+^{(n)}
  }{c_0}{a_0}{b_0}u\Km {K_-^{(n)}}{c_N}{a_N}{b_N}u\times  \no \\
  \prod_{k=0}^{N-1}\biggl[\wt
    {W_{(n\times m)}}{b_k}{b_{k+1}}{c_{k+1}}{c_k}{u}\wt
 {W_{(m\times n)}}{c_{k+1}}{a_{k+1}}{a_k}{c_k}{u+n\lambda-\lambda}
 \biggl].  \label{fopenT}
\ee
They form fusion hierarchies of commuting families of
transfer matrices
\be
\left[\mbox{\boldmath $T$}^{(m,n)}(u)\; ,
 \; \mbox{\boldmath $T$}^{(m,n^\prime)}(v)\;\right]=0\;,
 \label{fTT}
\ee
where the fusion level $m$ labels the different families and
fusion level $m=n=1$ is the unfused transfer matrix.
The fused face weights vanish when the fusion level $m$ is greater than
$L-1$ only for the ABF model \cite{ZhPe:94}. So we have $L-1$ families
in total for the ABF model while for the SOS model we have an
infinite number of families. In each family the
corresponding transfer matrices are related to each other and satisfy
a group of functional relations. To see this let us define
\be
\mbox{\boldmath $T$}^{(n)}_k&=&{\mbox{\boldmath $T$}}^{(m,n)}(u+k\lambda)
            \hspace{0.5cm} \mbox{$m=1,2,\cdots,L-1$}           \no \\
\mbox{\boldmath $T$}^{(n)}&=&0   \hspace{0.5cm} \mbox{if $n<0$
        or $m<0$}\no \\
\mbox{\boldmath $T$}^{(0)}&=&{\bf I}                                  \no
\ee
Then the functional relations read
\be
&&\hspace{0.5cm}\mbox{\boldmath $T$}^{(n)}_0\mbox{\boldmath $T$}^{(1)}_n=
   \mbox{\boldmath $T$}^{(n+1)}_0
  + f^m_{n-1}\mbox{\boldmath $T$}^{(n-1)}_0
     \hs{0.5}n\ge 0\label{fr}
\ee
with no closure condition ($n,m=1,2,\cdots$) for the SOS model but with
\be
\mbox{\boldmath $T$}^{(m,n)}(u)=0 \hspace{0.5cm} \mbox{if $n>L-1$}
\ee
as the closure condition ($m=1,2,\cdots,L-1$) for the ABF model.
The matrix function
$f^m_n(u)$ is related to the anti-symmetric fusion and is
dependent on both the bulk and boundary face weights.

The functional relations can be proved in a similar manner to
\cite{BOP:95,Zhou:95b}. We find the matrix function $f^m_n(u)$ to be given by
\be
f^m_n(u)&=&f^m(u+n\lambda)       \\
f^m(u)&=& {\omega^-(u)\omega^+(u)\over \rho(2u)}
      \prod_{j=0}^{m-1}[\rho(u-j\lambda)\rho(u+j\lambda)]^{N}  ,
\ee
where the boundaries contribute the factors $\omega^-(u)$
and $\omega^+(u)$, which are diagonal matrices labelled by the heights
along the right ($r$) and left ($l$) boundaries,
\be
\omega^-_{r,r^\prime}(u)=\sum_{a,b}\sqrt{\vartheta_1(w_b)\over
  \vartheta_1(w_{r-1})}\wt W{r-1}abr{2u+\lambda}\Km
  {K_-}bra{u+\lambda}\Km {K_-}{r\!-\!1}aru \delta_{r,r^\prime} &&\\
\omega^+_{l,l^\prime}(u)= \sum_{a,d}\!
  \sqrt{\vartheta_1^2(w_l)\vartheta_1^2(w_a)
    \over \vartheta_1^3(w_{l-1})\vartheta_1(w_d)}
  \wt Wda{l\!-\!1}l{\lambda\!-\!2u}\Km
  {K_+}{l\!-\!1}la{u\!+\!\lambda}\Km {K_+} dalu  \delta_{l,l^\prime} &&
\ee
Applying the boundary crossing symmetry to the surface face weights we have
\be
\omega^-_{r,r^\prime}(u)&=&{\vartheta_1({2\lambda+2u})
  \over\vartheta_1({\lambda})}\sum_{a} \Km {K}{r\!-\!1}ra{-u;\xi_-}
  \Km {K}{r\!-\!1}ar{u;\xi_-}\delta_{r,r^\prime}.
\ee
Similarly $\omega^+(u)$ is given by $\omega^-(u)$ with $u\to -u$,
$\xi_-\to\xi_+$ and $r\to l$. According to the spirit presented in
\cite{Zhou:95b,BaZh:95} the crossing unitarity condition
\be
T(u)T(u+\lambda)=f^1(u)\displaystyle{\rho(2u)\vartheta_1^2({\lambda})
/\vartheta_1^2({2\lambda})}  \label{inv-E}
\ee
determines both the bulk and surface free energies, where
$\rho(2u)\vartheta_1^2({\lambda})/\vartheta_1^2({2\lambda})$
has been used for the normalisation of the surface free energy.

\subsection{Surface critical phenomena}
\label{sec5}\setcounter{equation}{0}

The surface free energy can be obtained by applying the
``inversion relation trick'',  which is known to give the correct
bulk free energy of the eight-vertex model \cite{Baxter}.
The bulk and surface free energies must both satisfy the
unitarity relation (\ref{inv-E}) and can be
 separated from one another. In the following we consider only
diagonal $K$-matrices.

Let $\Lambda(u)=\Lambda_b(u)\Lambda_s(u)$ be the eigenvalues
of the transfer matrix $T(u)$. Define $\Lambda_b = \kappa_b^{2 N}$
and $\Lambda_s = \kappa_s$, then the free energies are defined by
$f_b(u)=-\log \kappa_b(u)$ and $f_s(u) = -\log \kappa_s(u)$.
For the bulk
\be
\kappa_b(u)\kappa_b(u+\lambda)&=&{\theta_1(\lambda-u)\theta_1(\lambda+u)
    \over \theta_1(\lambda)\theta_1(\lambda)} \label{inv-b34}
\ee
while for the surface
\be
&& \hspace{-1.2cm}
\kappa_s(u)\kappa_s(u+\lambda)=
  {\vartheta_1({2\lambda+2u})\vartheta_1({2\lambda-2u})
  \over\vartheta_1^2({2\lambda})} \\
&&\hspace{-0.5cm}\times  \Km {K}{r\!-\!1}rr{-u;\xi_-}
   \Km {K}{r\!-\!1}rr{u;\xi_-}
  \Km {K}{l\!-\!1}ll{-u;\xi_+}
   \Km {K}{l\!-\!1}ll{u;\xi_+}.  \no
\ee
At this point we need to distinguish between the unrotated and rotated
lattices. Naturally the bulk free energy remains unchanged in either
geometry.
For the rotated lattice in Fig.2 the boundary face weights
($K$-matrices) are gone, so the surface free energy is simply
determined by
\be
\kappa_s(u)\kappa_s(u+\lambda)&=&
  {\vartheta_1({2\lambda+2u})\vartheta_1({2\lambda-2u})
  \over\vartheta_1^2({2\lambda})}\;. \label{uni}
\ee

To obtain the surface free energy we consider the regime
$0<u<\lambda$ with $0 < p<1$, where
$p=e^{-\epsilon} \rightarrow 0$ at criticality.
To demonstrate the procedure, we first recall the
derivation of the bulk free energy.
We introduce the new variables
\be
x=e^{-4\pi\lambda/\epsilon},\hs{0.3}
w=e^{-4\pi u/\epsilon},\hs{0.3}
q = e^{-2\pi^2/\epsilon}.
\ee
Up to a harmless prefactor, which we disregard in the following, the
necessary conjugate modulus transformation of the theta function is
\be
\vartheta_1(u,e^{-\epsilon}) \sim
     E\left(e^{-4\pi u/\epsilon},
        e^{-4\pi^2/\epsilon}\right)
\ee
where
\be
E(z,x)=\prod_{n=1}^\infty(1-x^{n-1}z)(1-x^{n}z^{-1})(1-x^n).
\ee

The argument is to suppose that $\kappa_b(w)$ is analytic and nonzero
in the annulus $x\le w\le 1$ and perform the Laurent expansion
$\log\kappa_b(w)=\sum_{n=-\infty}^{\infty} c_n w^n$.
Then inserting this into the logarithm of both sides of (\ref{inv-b34})
and equating coefficients of powers of $w$ gives \cite{Baxter}
\be
f_b(u,p)=-2\sum_{n=1}^{\infty}
         {\sinh{[2\pi u n/\epsilon]}
     \sinh{\left[2\pi(\lambda- u) n/\epsilon\right]}
     \cosh{\left[2\pi(\pi-2\lambda) n/\epsilon\right]} \over  n
     \sinh{\left[2\pi^2 n/\epsilon\right]}
     \cosh{\left[2\pi\lambda n/\epsilon\right]} }\;. \label{bfree}
\ee
Similarly, Laurent expanding $\log\kappa_s(w)=
 \sum_{n=-\infty}^{\infty} c_n w^n$ and solving the
crossing unitarity relation (\ref{uni}) under the same
analyticity assumptions as for the bulk case gives
\be
f_s(u,p)
=-2\sum_{n=1}^{\infty}
       {\sinh{[4\pi u n/\epsilon]}
     \sinh{\left[4\pi(\lambda- u) n/\epsilon\right]}
     \cosh{\left[2\pi(\pi-4\lambda) n/\epsilon\right]} \over  n
     \sinh{\left[2\pi^2 n/\epsilon\right]}
     \cosh{\left[4\pi\lambda n/\epsilon\right]} }\;. \label{ks}
\ee

The model with $\lambda={1\over L+1}$ in the regime
$0<u<\lambda$ and $0<p<1$ is the $A_L$ ABF RSOS model
in regime I$\!$I$\!$I \cite{ABF:84}.
By making use of the Poisson summation formula,
the sum in (\ref{bfree}) can be rewritten in terms of $p$.
As $p\rightarrow 0$ the singular part of the bulk free
energy for $L$ odd scales as
\be
f_b \sim p^{2-\alpha_b} \log p \quad \mbox{with} \quad
\alpha_b= 2-{\pi\over 2\lambda}
\ee
with no singular contribution for $L$ even \cite{ABF:84}.

To extract the surface specific heat exponent $\alpha_s$, we follow
\cite{bin,ws} and define the local internal energy $e_s$ and
the surface specific heat $C_s$ in the surface layer,
\be
e_s(p)\sim {\partial f_s(w,p)\over\partial p},\hs{1}
 C_s \sim {\partial e_s\over\partial p}\;.
\ee
For the fixed boundary conditions we
drop the correction energy $e_1$ to $e_s(p)$ \cite{ws}.
 From (\ref{ks}) we find that as $p\to 0$
\be
f_s\sim p^{2-\alpha_s} \quad \mbox{with} \quad
\alpha_s=2-{\pi\over 4\lambda}.
\ee
There is a $\log p$ correction factor for $L$ odd.
The surface specific heat exponent for the $A_L$ ABF model follows as
\be
\alpha_s=2-\mbox{\small $\frac{L+1}{4}$}.
\ee
Using the known scaling relation $\alpha_s = \alpha_b + \nu$,
which has been explicitly confirmed for the eight-vertex model
\cite{BaZh:95}, we expect $\nu=(L+1)/4$ for the correlation length
exponent of the $A_L$ ABF  model with odd $L$.

Although the surface free energy on the original unrotated lattice
depends on the explicit form of the
boundary face weights,
they play no role in the dominant critical singularity
of the internal energy $e_s(p)$. This is seen also in the
study of the eight-vertex model \cite{BaZh:95}.
However, they do effect the correction energy $e_1$ to $e_s(p)$.
We therefore expect to see (presumably) height-dependent
surface free energies and thus other surface critical exponents.

\vskip 0.5cm
The authors thank Professor R. J. Baxter and Mr Vlad Fridkin
for helpful comments. This work has been supported by the Australian
Research Council.


\clearpage
\begin{thebibliography}{99}
\def\AP#1#2#3{\newblock{\sl Ann. Phys.} {\bf #1} (#2) #3}
\def\CMP#1#2#3{\newblock{\sl Commun. Math. Phys.} {\bf #1} (#2) #3}
\def\LMP#1#2#3{\newblock{\sl Lett. Math. Phys.} {\bf #1} (#2) #3}
\def\JSP#1#2#3{\newblock{\sl J. Stat. Phys.} {\bf #1} (#2) #3}
\def\JPA#1#2#3{\newblock{\sl J. Phys.} {\bf A#1} (#2) #3}
\def\JMP#1#2#3{\newblock{\sl J. Math. Phys.} {\bf #1} (#2) #3}
\def\IJMP#1#2#3{\newblock{\sl Int. J. Mod. Phys.} {\bf #1} (#2) #3}
\def\MPLA#1#2#3{\newblock{\sl Mod. Phys. Lett.} {\bf A#1} (#2) #3}
\def\NPB#1#2#3{\newblock{\sl Nucl. Phys.} {\bf B#1} (#2) #3}
\def\PLA#1#2#3{\newblock{\sl Phys. Lett.} {\bf A#1} (#2) #3}
\def\PLB#1#2#3{\newblock{\sl Phys. Lett.} {\bf B#1} (#2) #3}
\def\PRL#1#2#3{\newblock{\sl Phys. Rev. Lett.} {\bf#1} (#2) #3}
\def\PR#1#2#3{\newblock{\sl Phys. Rev.} {\bf#1} (#2) #3}
\def\PTP#1#2#3{\newblock{\sl Prog. Theor. Phys. } {\bf#1} (#2) #3}
\def\LMP#1#2#3{\newblock{\sl Lett. Math. Phys.} {\bf#1} (#2) #3}
\def\PA#1#2#3{\newblock{\sl Physica A } {\bf#1} (#2) #3}
\def\SPJ#1#2#3{\newblock{\sl Sov. Phys. JETP. } {\bf#1} (#2) #3}


\bibitem{Sklyanin}E. K. Sklyanin, \JPA {21}{1988}{2375}.
\bibitem{MezNep} L. Mezincescu and R. I. Nepomechie,
``Lectures on Integrable Quantum Spin Chains,''
in {\em New Developments of Integrable Systems and Long-Ranged
Interaction Models}, ed. by M.-L. Ge and Y.-S. Wu,
World Scientific Publishing Co., 96 - 142 (1995).

\bibitem{YuBa:95} C. M. Yung and M. T. Batchelor, \NPB {435}{1995}{430}.
\bibitem{Kulish} P. P. Kulish, {\em Yang-Baxter equation and reflection
equations in integrable models}, hepth/9507070.
\bibitem{BOP:95}R. E. Behrend, P. A. Pearce and David L. O'Brien,
  {\em  Interaction-Round-a-Face
  models with fixed boundary conditions: the ABF fusion hierarchy,}
  hep-th/9507118.
\bibitem{GhZa:94}S. Ghoshal and A. Zamolodchikov,
       \IJMP {A 21}{1994}{3841}.
\bibitem{Baxter:73}R. J. Baxter, \AP {76}{1973}{25}.
\bibitem{Baxter:72}R. J. Baxter, \AP {70}{1972}{193}.
\bibitem{ABF:84}G.~E.~Andrews, R.~J.~Baxter and P.~J.~Forrester,
     \JSP {35}{1984}{193}.
\bibitem{Zhou:95b}Y. K. Zhou, {\em Row transfer matrix functional
relations for Baxter's eight vertex model with open boundaries
   via more general reflection matrices,} hep-th/9510095,
     to be published in  {\sl Nucl. Phys.} B (1995).
\bibitem{AK:95} C. Ahn and W. M. Koo, {\em Boundary Yang-Baxter equation
in the RSOS representation}, hepth/9508080.

\bibitem{SaBa:89}H. Saleur and M. Bauer, \NPB {320}{1989}{591}.
\bibitem{Cardy:89}J. L. Cardy, \NPB {324}{1989}{581}.
\bibitem{DeDe:92}C. Destri and H. J. de Vega, \NPB {374}{1992}{692}.

\bibitem{DJKMO:88} E. Date, M. Jimbo, A. Kuniba, T. Miwa and M. Okado,
  {\sl Adv. Stud. Pure Math.,} {\bf 16}(1988) 17.
\bibitem{ZhPe:94}Y.~K.~Zhou and P.~A.~Pearce,
  \newblock \IJMP {B8}{1994}{3531}.

\bibitem{MeNe:92}L. Mezincescu and R. I. Nepomechie,
          \JPA {25}{1992}{2533}.

\bibitem{Zhou:95a}Y. K. Zhou, \NPB {453}{1995}{619}.
\bibitem{BaZh:95}M. T. Batchelor and Y. K. Zhou,
  {\em Surface Critical Phenomena and Scaling in the Eight-Vertex Model},
 cond-mat/9510152, to be published in  {\sl Phys. Rev. Lett.}
\bibitem{Baxter}R. J. Baxter, 
 ``{\em Exactly Solved Models in Statistical
  Mechanics"}, Academic Press, London, 1982.

\bibitem{bin} K. Binder, in {\it Phase Transitions and Critical Phenomena},
edited by C. Domb and J. L. Lebowitz, (Academic, London, 1983), Vol. 8, p 1.
\bibitem{ws}M. Wortis and N. M. ${\check {\rm S}}$vraki{$\acute {\rm c}$},
   {\sl IEEE Tran.   Mag.} {\bf 18}, 721 (1982).

\end{thebibliography}
\end{document}